\documentclass{eptcs} 
\providecommand{\event}{Axel Legay (Ed.): International Workshop on Verification of Infinite-State Systems (INFINITY)} 
\providecommand{\volume}{13}   
\providecommand{\anno}{2009}   
\providecommand{\firstpage}{1} 
\providecommand{\eid}{1}       
\usepackage{latexsym}
\usepackage{amsmath}
\usepackage{amsfonts}
\usepackage{amssymb}
\usepackage{amsthm}
\usepackage{gastex}
\usepackage{mathrsfs}
\usepackage{times}
\usepackage{subfig}
\usepackage{graphicx}
\usepackage{multirow}
\usepackage{rotating}

\newtheorem{theorem}{Theorem}[section]
\newtheorem{lemma}[theorem]{Lemma}
\newtheorem{proposition}[theorem]{Proposition}

\newtheorem{definition}[theorem]{Definition}

\newcommand{\N} {{\mathbb N}}
\newcommand{\Z} {{\mathbb Z}}
\newcommand{\R} {{\mathbb R}}
\newcommand{\Q} {{\mathbb Q}}

\newcommand{\M} {{\cal M}}

\newcommand{\g} {{\bf g}}

\newcommand{\x} {{\bf x}}

\title{Dense-choice Counter Machines revisited}

\author{Florent Bouchy, Alain Finkel
	\institute{LSV, ENS Cachan, CNRS}
	\email{$\lbrace$bouchy,finkel$\rbrace$@lsv.ens-cachan.fr}
	\and
	Pierluigi San Pietro
	\institute{Dipartimento di Elettronica e Informazione, Politecnico di Milano} 
	\email{pierluigi.sanpietro@polimi.it}
}

\def\titlerunning{Dense-choice Counter Machines revisited}
\def\authorrunning{F. Bouchy, A. Finkel, P. San Pietro}

\begin{document}

\maketitle

\begin{abstract}
This paper clarifies the picture about Dense-choice Counter Machines, which have been less studied than (discrete) Counter Machines. We revisit the definition of "Dense Counter Machines" so that it now extends (discrete) Counter Machines, and we provide new undecidability and decidability results. Using the first-order additive mixed theory of reals and integers, we give a logical characterization of the sets of configurations reachable by reversal-bounded Dense-choice Counter Machines. 
\end{abstract}


\section{Introduction}

\emph{Discrete} (i.e. integer-valued) Counter Machines have been
well-studied and still receive a lot
of attention. We can mention Minsky Machines \cite{Minsky67}, different kinds of counter systems (e.g., \cite{BFLP-sttt08,DDFG-atva06},
which are Minsky Machines using affine functions instead of increment/decrements and zero-tests, or \cite{CJ98,iosif06}), 
Petri nets (or equivalently, VASS), and their many extensions. 

There are also extensions of discrete counter systems to real-valued systems, 
called \emph{hybrid} systems, such as linear hybrid automata, real-counter systems \cite{RealCounter}, or dense counter
systems \cite{DenseCounter}.
Another subclass of hybrid systems is
the well-known decidable model of Timed Automata \cite{alur94}, which has been linked to special classes of counter systems in \cite{CJ98} and \cite{iosif06}.
Recently, some
connections between Timed Automata and timed Petri nets have been made
\cite{RHS-atva07,BHR-ietc07}. An extension of counter systems to timed counter systems
has been defined and studied in \cite{BFS-infinity08}.

Linear hybrid automata \cite{LHA}, as well as
Timed Automata extended with only one stopwatch, are already 
undecidable. Other subclasses of hybrid systems, like hybrid Petri nets (stochastic Petri nets,
continuous Petri nets, differential Petri nets, timed Petri nets) are
dense extensions of Petri nets, but they have not the same semantics and
their comparison is not always easy or feasible (see \cite{dotoli08} for a recent
survey).\\

On the other side, from our point of view, the natural extension of (discrete)
counter systems to dense counter systems is quite recent; to the best of our
knowledge, the first paper which introduces Dense
Counter Machines (DCM) as a natural generalization of Counter Machines (CM) is \cite{DenseCounter}. Their
Dense Counter Machine allows
incrementing/decrementing each counter by a non-deterministically-chosen real
$\delta$ between $0$ and $1$. The motivation of this extension is 
to model hybrid systems where a nondeterministic choice can
be made (see for example the argumentation about the dense
producer/consumer in \cite{DenseCounter}, which neither Timed Automata nor hybrid automata can model in an easy way).
However, what can we earn from extending CM (which have the total expression
power of computability) into DCM ? Non-trivial problems will
remain, of course, undecidable. The direction followed by \cite{DenseCounter} is to
find subclasses of DCM for which the binary reachability is still computable, such as reversal-bounded DCM.

\paragraph{\bf Our contributions.}
We give a general definition of Counter Systems containing all
the variations of counters (discrete, dense-choice, purely dense-choice, etc.). We then revisit the definition of "Dense Counter Machines" \cite{DenseCounter} into 
\emph{Dense-choice Counter Machines} (shortly, also, \emph{DCM}) so that it is now simpler, more precise and formal, and also more clearly
understandable as a natural extension of Minsky Machines. 
A DCM
is a finite-state machine augmented with
dense-choice counters, which can assume only non-negative
real values. At each step, every dense-choice counter
can be incremented/decremented by $0$, $1$,
or by a non-deterministically-chosen $\delta$, $0<\delta<1$ (which is supposed different at each step, since the choice is random). 
We assume w.l.o.g. that
for a given step,  the {\em same} $\delta$ is used for all the counters.
This $\delta$ increment/decrement is the essential difference
between dense-choice and discrete counters. A DCM can also test a
counter $x_i$ against $0$ (either $x_i=0$ or $x_i>0$). 

Since dense-choice counters are
(trivially) more general than discrete
counters, we also study the model of \emph{purely-DCM}, i.e. DCM in which 
counters lose the ability to increment/decrement by $1$.
We show that the
restriction to {\em bounded} purely-DCM (i.e. there exists a constant bound $b$ such
that each counter is bounded by $b$) still produces an undecidable
control-state reachability problem (even with four $1$-bounded
purely dense-choice counters).\\

We then consider an effective (i.e. whose binary reachability is computable)
class of DCM: reversal-bounded DCM \cite{DenseCounter}. In order to
model hybrid systems more easily, we wish to
introduce the ability for a counter to be tested against an integer $k$
(instead of $0$): this is an easy, common extension 
for Minsky Machines and for Petri nets, but it produces new technical problems
for reversal-bounded DCM. One of the reasons is that the usual simulation of a $k$-test 
(i.e., several decrements and $0$-tests, followed by increments restoring the original counter value) does not preserve reversal-boundedness. 
We actually show that reversal-bounded DCM with $k$-tests are equivalent
to reversal-bounded DCM, using a long and technical proof. This allows us to obtain as a corollary that the reachability
relation of a DCM with one free counter and a finite number of 
reversal-bounded $k$-testable counters is still effectively 
definable by a mixed formula (this extends a previous result of \cite{DenseCounter}).\\

Using the first-order additive mixed theory of
reals and integers, $\mathrm{FO}(\R,\Z,+,<)$, we give a logical characterization of the sets of
configurations reachable by reversal-bounded DCM. We
prove that any mixed formula is the reachability relation of a
reversal-bounded DCM. This completes the initial result stating 
that the reachability relation of a reversal-bounded DCM is definable by
a mixed formula.

\section{Dense-choice Counter Machines}

\paragraph{\bf Notations.} 
We use $\R$ to denote the set of real numbers, $\R_+$ the set of non-negative real numbers, $\Q_+$ the set of non-negative rational numbers, $\Z$ the set of integers, and $\N$ the naturals. Capital letters (eg. $X$) denote sets, and small letters (eg. $x$) denote elements of sets. Bold-faced symbols (eg. $\x$) denote vectors, and subscripted symbols (eg. $x_i$) denote components of vectors. Sometimes, for the sake of readability, we use $x$ instead of $x_i$ (without real ambiguity). Throughout this paper, $n \in \N$ is the number of counters.

\subsection{Extending Minsky Machines}
\label{sec:Minsky}
In this section, we motivate the use of Dense-choice Counter Machines, by arguing about possible ways to extend Minsky Machines \cite{Minsky67}. Minsky Machines are indeed the most elementary definition of Counter Systems that we will consider here, and probably the most known. A Minsky Machine has a finite set of control states, and operates transitions between them, by executing instructions on a finite set of integer-valued variables (the counters). Its possible instructions are (1) increment a counter value by $1$, (2) test if a counter value is $0$, and (3) if a counter value is greater than $0$ then decrement it by $1$.\\

Let $\mathscr{L}$ be a given logic, such as the Presburger logic $\mathrm{FO}(\N,+,=)$, $\mathrm{FO}(\R,\Z,+,<)$, etc. A formula $\mathcal{F}(\x,\x')$ of $\mathscr{L}$, with $2n$ free variables, is interpreted as the tranformation of counter values $\x$ into $\x'$: it defines the counter values \textit{before} and \textit{after} the firing of a transition labelled by $\mathcal{F}(\x,\x')$. Throughout this paper, we will use several different classes of counter machines, each one based on the generic Definition~\ref{def:CM}. They all use a finite labelling alphabet $\Sigma \subseteq \mathscr{L}$, defining instructions on a vector $\x = (x_1,x_2,\ldots,x_n)$. The way the alphabet $\Sigma$ is defined is what makes the difference between various Counter System classes.
\medskip

\begin{definition} \label{def:CM}
A \emph{Counter System} (CS for short) is a tuple $\M = \langle S,T\rangle$ such that $S$ is a finite set of control states, and $T \subseteq S \times \Sigma \times S$ is a finite set of transitions.
\end{definition}

\medskip
Remark that a Minsky Machine is a CS in which formulas of $\Sigma$ are of the form $ (x'= x+1)$, $(x'=x=0)$, $(x>0 \land x'=x-1), $ or $true$ ($x$ being a component of $\x$, i.e., a counter $x_i$). Although the reachability problem is undecidable for Minsky Machines (with at least two counters), we would like to extend them for two reasons. First, if a Minsky Machine is reversal-bounded, then its reachability relation is computable; thus, we would like to use a more powerful model than reversal-bounded Minsky Machines, which remains decidable. This first point will be detailed in sections \ref{sec:DCM} and \ref{sec:RB}. Second, Minsky Machines are very basic and not practical to use for modelling or expressing high-level properties. For that matter, we add the possibility to use real-valued counters, and to non-deterministically choose the value of an increment/decrement for each transition. In the remainder of this section, we discuss these two extensions.\\

In order to get real-valued counters, we define Dense Minsky Machines, which are CS whose $\Sigma$ is composed of formulas of the form $(x'= x+r)$, $(x'=x=0)$, $((x-r>0 \lor x-r=0) \land x'=x-r)$, or $true$, with a given finite set of values\footnote{Note that here, we take these values in $\Q_+$ 
because the important properties are (1) density and (2) an effective representation of any rational number (this is not the case for reals, in general).}
$r \in \Q_+$. Like in Minsky Machines, the initial counter values are always $0$. This first extension is not really more powerful, since it can be simulated by a Minsky Machine:
\begin{proposition}\label{prp:minsky}
Minsky Machines and Dense Minsky Machines are bisimilar.
\end{proposition}
\begin{proof}
One way is obvious, by taking $r=1$. The other way is a little more elaborate, but remains easy. We just have to simulate every Dense Minsky Machine instruction with a Minsky machine. There are four instructions, and two of them are obviously the same: $true$ and $x'=x=0$. For the two other instructions, $x'=x+r$ and $(x-r>0 \lor x-r=0) \land x'=x-r$, we just have to encode $r$ by an integer. Each increment/decrement $r\in \Q_+$ can be written as $\frac{p}{q}$, with $p,q \in \N$. Then, since we know all the possible $r$ in advance, we can compute for each $r$ a $q' \in \N$ such that $r=\frac{pq'}{q_{lcm}}$, where $q_{lcm}$ is the least common multiple of all $q$. Thus, each $r$ can be represented by a non-negative integer $r'=pq'$, and the new counter values will all be multiplied by the same factor $q_{lcm}$. Using this simple encoding, we can simulate an instruction $x'=x+r$ by a sequence of $r'$ instructions $x'=x+1$. Likewise, $(x-r>0 \lor x-r>0) \land x'=x-r$ can be simulated by a sequence of $r'$ instructions $x>0 \land x'=x-1$.
\end{proof}

Another way to extend Minsky Machines is to allow, on each transition, a non-deterministic choice of the increment/decrement. We call this extension a Dense-choice Minsky Machine, which is a CS whose $\Sigma$ contains formulas of the form $ (x'= x+1)$, $(x'=x=0)$, $(x>0 \land x'=x-1)$, $(x'=x+\Delta)$, $((x-\Delta >0 \lor x-\Delta =0) \land x'=x-\Delta)$, or $true$, where $\Delta$ symbolizes a non-deterministically-chosen value $\delta \in \R_+$. The choice is made each time a transition is fired, so that two consecutive transitions labelled by $x'=x+\Delta$ should generally have different values for $\Delta$: the choice is random, and we have no knowledge of the chosen value (although we could check it afterwards, using an additional counter and transition).

\medskip
Here, we consider runs of {\em finite} length only: we show that the $\delta$ value can be chosen in $]0,1[$ instead of $\R_+$:
\begin{proposition}
For finite-length runs, every Dense-choice Minsky Machine $M$ with $\delta \in \R_+$ can be simulated by a Dense-choice Minsky Machine $\tilde{M}$ with $\delta \in\ ]0,1[$.
\end{proposition}
\noindent\textit{Sketch of the proof.} Every run $r$ of such a machine $M$ can be simulated by a possibly longer run $\tilde{r}$ of $\tilde{M}$ 
whose $\delta$ increments are in the open interval $]0,1[\,$.
Without a formal proof and formal definitions of control state and configuration, we show how, for instance, a transition from control states $s$ to $s'$, 
labelled by $(x_1' = x_1+\Delta)$, $(x_2'=x_2+\Delta)$, and $((x_3-\Delta >0 \lor x_3-\Delta =0) \land x_3'=x_3-\Delta)$, 
can be simulated in $\tilde{M}$ (each $\Delta$ being replaced at each firing of a transition by a $\delta \in \R_+$).

First, $\tilde{M}$ has a transition from $s$ to a new control state $s''$, 
labelled by the same formulas, but with $0<\delta<1$.
Second, in $\tilde{M}$ there is also a transition from $s''$ to $s''$ itself, 
again labelled by the same formulas, with $0<\delta<1$.
Third, in $\tilde{M}$ there is a transition  from $s''$ leading to $s'$  labelled by $x'_1=x_1$, $x'_2=x_2$, $x'_3=x_3$.
Hence, a configuration $c'$ with control state $s'$ and counter values $(x'_1, x'_2, x'_3)\in \R_{+}^3$ is reachable in $M$ from a configuration $c$ in control state $s$ and counter values 
$(x_1,x_2,x_3) \in \R_{+}^3$  iff $c'$ is reachable in $\tilde{M}$ from $c$. \qed \\

Therefore, there is no loss in generality in assuming that each increment is in the interval $]0,1[$, at least as long as finite runs are considered.
Instead, a bounded increment can give a finer degree of control on counters. In fact, in many physical systems, physical variables are 
actually bounded (e.g., a water level in a reservoir, which is a non-negative real value that cannot exceed the height of the reservoir).  
It seems difficult to model or check this kind of behaviour with a CS where increments are unbounded reals. 

Finally, we notice that allowing increments in the interval $]0,q[$, with a fixed $q\in \N$, does not give any gain in expressivity with respect to the case of $q=1$.
For instance, to increment a counter $x$ by any value $\delta$ with $0<\delta<q$, it is enough to apply, in a Dense-choice Minsky Machine with non-determistic increments in $]0,1[\,$, 
a sequence of exactly $q$ transitions (this is possible, since $q$ is fixed), each of the form $x'=x+\delta$, for $0<\delta<1$.

In the next section, we generalize and formalize the definition of Dense-choice Minsky Machine that we just motivated.

\paragraph{Example: producer-consumer system.} As a simple example of application of a machine with real-valued counters, consider the following version of a traditional
producer-consumer system, described in \cite{DenseCounter}.
A system may be in one of three states: {\em produce},
{\em consume} or {\em idle}. When in state {\em produce}, a
resource is created, which may be stored and later used while
in state {\em consume}. The resource is a real number,
representing an available amount of a physical quantity, such as fuel or water.
Production may be stored, and used up much later (or not used at all).
This system may be easily modeled by a finite state machine with
one dense-choice counter, which is shown below, where the resource is added when produced
or substracted when consumed. 

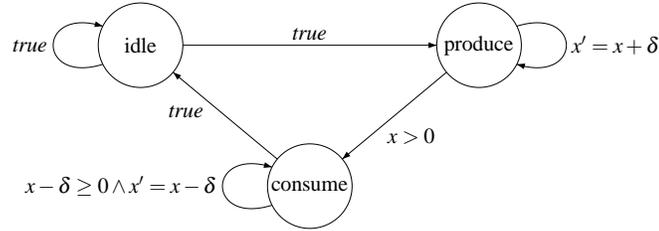
\begin{center}
\scalebox{0.75}{
  \begin{picture}(100,40)(-15,0)
\label{fig:example}
    \gasset{Nw=15,Nh=15, Nmr=10}
    \node(A)(5,30){idle}
    \node(B)(65,30){produce}
    \node(C)(35,5){consume}
    \drawedge(A,B){\emph{true}}
    \drawedge(B,C){$x > 0$}
    \drawedge[ELpos=65](C,A){\emph{true}}
	\drawloop[loopangle=180](A){\emph{true}}
	\drawloop[loopangle=0](B){$x' = x+\delta$}
	\drawloop[loopangle=180](C){$x-\delta \geq 0 \land x' = x-\delta$}
  \end{picture}}
\end{center}

Using a real-valued counter, there is an underlying assumption
that a continuous variable, such as this resource,
changes in discrete steps only; however,
this is acceptable in many cases where a variable actually changes continuously,
since the increments/decrements may be arbitrarily small.
Since the counter may never decrease below zero,
the specified system implements the physical
constraint  that consumption must never exceed production. 
More complex constraints are decidable, for instance if expressed by linear constraints on counter values. 
An example of a decidable query is whether total production never exceeds twice the consumption.


\subsection{Definitions and properties of Dense-choice Counter Machines}
\label{sec:DCM}

Let $\x$ be a vector of $n$ variables, called \emph{dense-choice counters} (or simply \emph{counters} if not specified otherwise). 
Dense-choice counters were called ``dense counters" in \cite{DenseCounter}. A \emph{counter valuation} is a function giving, for any $x_i \in \x$, a value in $\R_+$. In this paper, we write $x_i$ (or $\x$) to denote both variable(s) and the image of valuation(s), since there is no ambiguity and the meaning is obvious.
Let $G = \lbrace (x=0), (x>0), true \rbrace$ be the set of guards. We say that a counter valuation $\x$ satisfies a guard $\g\in G^n$ with the usual meaning, denoted by $\x  \models \g$; for example, if $n=3$ and $\g=(true, x_2>0, x_3=0)$, then $(6,2,0)\models \g$ but $(6,2,1)\not\models \g$. Let $A = \lbrace 1, \Delta \rbrace$ be the set of actions; intuitively, $1$ stands for an integer increment/decrement, and $\Delta$ stands for a non-deterministically-chosen real increment/decrement.

\medskip
\begin{definition}
A \emph{Dense-choice Counter Machine} (shortly, a DCM) with $n>0$ counters is a tuple $\M = \langle S, T \rangle$ where:
\begin{itemize}
\item $S$ is a finite set of control states, with a state $s_{fin}\in S$ called the {\em final state} of $\M$;
\item $T \subseteq S \times \Sigma \times S$ is a finite set of transitions, with $\Sigma = (G \times \Z \times A)^n$
\end{itemize}
\end{definition}

\medskip
Intuitively, the integer component $\boldsymbol{\lambda} \in \Z^n$ of $\Sigma$ is a factor determining whether the transition is incrementing or decrementing a counter, and of which value. Meanwhile, the action $\mathbf{a} \in A^n$ determines whether the increment or decrement is a real or integral value. 
For the sake of clarity, we sometimes write transitions as $x>0 \land x:=x+3\delta$, meaning that the guard on counter $x_i$ is $g_i = (x>0)$, its factor is $\lambda_i = 3$, and its action is $a_i = \Delta$.

Notice that our transitions are equivalent to those of \cite{DenseCounter,RealCounter}, in which the authors used the notion of \emph{modes}. 
The modes \texttt{\small{stay}}, \texttt{\small{unit increment}}, \texttt{\small{unit decrement}}, \texttt{\small{fractional increment}}, and \texttt{\small{fractional decrement}} are here respectively represented by the cases $(\lambda_i=0)$, $(\lambda_i>0 \land a_i=1)$, $(\lambda_i<0 \land a_i=1)$, $(\lambda_i>0 \land a_i=\Delta)$, and $(\lambda_i<0 \land a_i=\Delta)$.
Also notice that transitions where $\lambda \in \lbrace+1,-1\rbrace^n$ are just a special case, and that they can simulate a linear combination of the form $x'_i = x_i + \sum_{j=1}^m \lambda_j \delta_j$, for a given $m$ and a vector of different $\delta_j$ values in $]0,1[\,$.\\

As usually done in verification, to interpret a DCM, we specify an initial valuation to each counter and an initial control state, and then we let the machine behave non-deterministically. The behaviour of a DCM mainly consists in choosing a transition whose guard $\g \in G^n$ is satisfied by the current counter valuations, and to update these valuations (and, of course, to go to the new control state).

\begin{definition}
The semantics of a DCM $\M = \langle S,T \rangle$ is given by a transition system $TS(\M) = \langle C,\rightarrow \rangle$ where:
\begin{itemize}
\item $C = S\times \R_+^n$ is the set of configurations
\item $\rightarrow \subseteq C \times \Sigma \times C$ is the set of transitions, defined by:\\
$(s,\x) \stackrel{\g, \boldsymbol{\lambda}, \mathbf{a}}{\longrightarrow} (s',\x') $ if and only if $(s, (\g, \boldsymbol{\lambda}, \mathbf{a}), s') \in T$ 
and $\exists\delta \in \R$ such that:\\ $0 < \delta <1 \land \x \models \g \land \x' = \x + \boldsymbol{\lambda} \mathbf{u}$, with $\mathbf{u} = \mathbf{a}[\Delta \leftarrow \delta]$

\end{itemize}
\end{definition}

For a DCM $\M = \langle S,T \rangle$ and its transition system $TS(\M) = \langle C,\rightarrow \rangle$, the reachability relation $\leadsto_\M$ is the reflexive and transitive closure ${\rightarrow^*}$; 
when the context is clear, we drop the subscript ${}_\M$. A {\em run} of $\M$ is a sequence $(s^0,\x^0)\rightarrow(s^1,\x^1)\rightarrow \ldots \rightarrow(s^l,\x^l)$, of length $l\ge 0$. 
Because of the inherent non-determinism of a DCM, we are interested only in runs ending in final state $s_{fin}\in S$. 
Formally, a run of $\M$ is \emph{accepting} if it is of the form $(s,\x)\rightarrow^* (s_{fin},\x')$, for some $s\in S$ and $\x,\x' \in \R_+^n$; $\M$ is also said to {\em accept}.
A DCM $\M$ is said to {\em reject} (or {\em crash}) during a run $(s,\x)\rightarrow^* (s',\x')$, if $s'\in S$ is a non-final sink state (hence, the run cannot be extended to be an accepting run).

The set of all pairs $\Bigl((s,\x),(s',\x')\Bigr)\in C \times C$ such that $(s,\x)\leadsto (s',\x')$ is called the {\em binary reachability relation} of $\M$; we sometimes call it {\em binary reachability} or {\em reachability relation}. 
The \emph{binary reachability problem} consists in computing\footnote{By "computation", we mean the existence of an algorithm which computes a formula (e.g. as a binary automaton). In general, such computation does not exist. } the binary reachability relation of a given DCM. 
An easier version is the \emph{control-state reachability problem} (shortly, \emph{state reachability problem}), which consists in deciding whether a given control state is reachable in some accepting run of a given DCM.

The example at the end of section \ref{sec:Minsky} is a DCM, if we remove the guard $x-\delta \geq 0$ (the machine crashes if the guard is not satisfied).\\

Although a DCM has only a restricted set of possible operations on counters, it can perform various higher-level macros, such as reset, copy, addition, substraction, comparison, etc. Here, we give the encodings of some of these macros, in order to be able to use them as shorthands in this paper.

We denote by \scalebox{0.8}{\begin{picture}(27,8)
      \gasset{Nw=5,Nh=5}
      \node(A)(3,1){$s$}
      \node(B)(24,1){$s'$}
      \drawedge(A,B){$\mathtt{reset}(x)$}
\end{picture}}, 
\scalebox{0.8}{\begin{picture}(27,8)
      \gasset{Nw=5,Nh=5}
      \node(A)(3,1){$s$}
      \node(B)(24,1){$s'$}
      \drawedge[ELdist=2](A,B){$\mathtt{copy}(x,y)$}
\end{picture},} 
\scalebox{0.8}{\begin{picture}(27,8)
      \gasset{Nw=5,Nh=5}
      \node(A)(3,1){$s$}
      \node(B)(24,1){$s'$}
      \drawedge(A,B){$\mathtt{add}(x,y)$}
\end{picture}}, and
\scalebox{0.8}{\begin{picture}(27,8)
      \gasset{Nw=5,Nh=5}
      \node(A)(3,1){$s$}
      \node(B)(24,1){$s'$}
      \drawedge[ELdist=2](A,B){$\mathtt{minus}(x,y)$}
\end{picture}} the DCM on Figures \ref{fig:reset}, \ref{fig:copy}, \ref{fig:add}, and \ref{fig:minus} (respectively).\\

\begin{figure}[h!]
\centering
\hfill
\subfloat[Reset of a counter]{\label{fig:reset}
\begin{picture}(30,15)(0,2)
      \gasset{Nw=5,Nh=5}
      \node(A)(5,5){$s$}
      \node(B)(22,5){$s'$}
      \drawedge(A,B){$x=0$}
	  \drawloop[loopdiam=4](A){$x:=x-\delta$}
\end{picture}}
\hfill
\subfloat[Copy of a counter into other ones]{\label{fig:copy}
    \begin{picture}(70,15)(0,2)
      \gasset{Nw=5,Nh=5}
      \node(A)(5,5){$s$}
      \node(B)(25,5){}
      \node(C)(45,5){}
      \node(D)(65,5){$s'$}
      \drawedge[ELdist=-4.4](A,B){\shortstack{$\mathtt{reset}(y)$\\$\mathtt{reset}(z)$}}
      \drawedge(B,C){$x=0$}
      \drawedge(C,D){$z=0$}
      \drawloop[loopangle=90,loopdiam=4](B){\shortstack{$x:= x-\delta$\\$y:= y+\delta$\\$z:= z+\delta$}}
      \drawloop[loopangle=90,loopdiam=4](C){\shortstack{$x:= x+\delta$\\$z:= z-\delta$}}
    \end{picture}}\\
\hfill
\subfloat[Addition of a counter into another one]{\label{fig:add}
\begin{picture}(55,20)(-15,2)
      \gasset{Nw=5,Nh=5}
      \node(A)(5,5){$s$}
      \node(B)(22,5){$s'$}
      \drawedge(A,B){$y=0$}
	  \drawloop[loopdiam=4](A){\shortstack{$x:=x+\delta$\\$y:=y-\delta$}}
\end{picture}}
\hfill
\subfloat[Substraction of a counter from another one]{\label{fig:minus}
\begin{picture}(60,20)(-16,2)
      \gasset{Nw=5,Nh=5}
      \node(A)(5,5){$s$}
      \node(B)(22,5){$s'$}
      \drawedge(A,B){$y=0$}
	  \drawloop[loopdiam=4](A){\shortstack{$x:=x-\delta$\\$y:=y-\delta$}}
\end{picture}}
\caption{Encodings of \texttt{reset}, \texttt{copy}, \texttt{add}, and \texttt{minus} operations}
	\label{fig:operations}
\end{figure}

\bigskip

Let $G'_k = \lbrace (x=i), (x<i), (x>i), true \rbrace_{i \in [0,k]}$, for a given $k \in \N$. A DCM whose set of guards is included in $G'_k$ is called a \emph{$k$-DCM}. The counters of a $k$-DCM are said \emph{$k$-testable}. Notice that a DCM is a $0$-DCM, and note that dense-choice counters are $0$-testable if not specified otherwise.

\begin{proposition} \label{prp:k-testable}
DCM can simulate $k$-DCM, for any given $k \in \N$.
\end{proposition}
\begin{proof}
There are three different kinds of additional tests: $x<k$, $x>k$, and $x=k$, for any given $k\in \N$. We show how to encode each of them with only $x=0$ and $x>0$ tests.

A test $x<k$, represented by a transition \begin{picture}(23,5)
      \gasset{Nw=5,Nh=5}
      \node(A)(3,1){$s$}
      \node(B)(20,1){$s'$}
      \drawedge(A,B){$x<k$}
    \end{picture}, can be simulated by the following encoding:\\
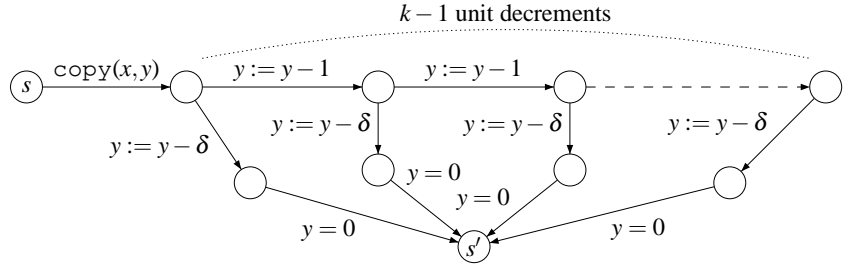
  \begin{center}
\scalebox{0.85}{
    \begin{picture}(110,40)(-10,0)
      \gasset{Nw=5,Nh=5}
      \node(Z)(-20,30){$s$}
      \node(A)(5,30){}
      \node(B)(35,30){}
      \node(C)(65,30){}
      \node(D)(105,30){}
      \node(E)(50,5){$s'$}
	  \node(AA)(15,15){}
	  \node(BB)(35,17){}
	  \node(CC)(65,17){}
	  \node(DD)(90,15){}
      \drawedge(Z,A){\texttt{copy}($x,y$)}
      \drawedge(A,B){$y := y-1$}
      \drawedge(B,C){$y := y-1$}
      \drawedge[dash={1.5}0](C,D){}
      \drawedge[ELpos=30,ELside=r](A,AA){$y := y-\delta$}
      \drawedge[ELpos=38,ELside=r](AA,E){$y=0$}
      \drawedge[ELpos=45,ELside=r](B,BB){$y := y-\delta$}
      \drawedge[ELpos=38,ELside=l](BB,E){$y=0$}
      \drawedge[ELpos=45,ELside=r](C,CC){$y := y-\delta$}
      \drawedge[ELpos=70,ELside=r](CC,E){$y=0$}
      \drawedge[ELpos=75,ELside=r](D,DD){$y := y-\delta$}
      \drawedge[ELpos=38,ELside=l](DD,E){$y=0$}
	\gasset{Nframe=n}
      \node(F)(5,34){}
      \node(G)(105,34){}
	  \drawedge[AHnb=0,ATnb=0,curvedepth=5,dash={0.2 0.5}0](F,G){$k-1$ unit decrements}
    \end{picture}}
  \end{center}

Note that to avoid modifying the value of $x$, we copy it into another counter $y$ by using the encoding of Fig.\ref{fig:copy}. 
In an easier way, a test $x>k$ (resp. $x=k$) can be simulated by a sequence of $k$ unit decrements followed by a test $x>0$ (resp. $x=0$). 
Finally, remark that the test $x<0$ can never be verified, since a counter cannot take negative values: instead, the machine would crash. 
\end{proof}

\begin{definition}
\label{def:purely}
Let $\M$ be a DCM. A counter $x_i$ of $\M$ is {\em purely dense-choice} if and only if $a_i=\Delta$ in every transition (i.e., it is never incremented/decremented by 1). Conversely, a counter $x_i$ is {\em (purely) discrete} if and only if $a_i=1$ in every transition (i.e., it is a classical discrete counter).
If $\M$ contains only purely dense-choice counters, it is called a \emph{purely-DCM}. If $\M$ contains only discrete counters, it is called a \emph{(discrete) Counter Machine (CM)}, as defined in \cite{ibarra-reversal-78}.
\end{definition}

\subsection{Reversal-Bounded DCM}
\label{sec:RB}

To extend the definition of reversal-boundedness from \cite{ibarra-reversal-78} to DCM, let $\M=\langle S,T \rangle$ be a DCM, $s,s'\in S$, and $r\in\N$. 
On a run from $s$ to $s'$, a counter $x_i$ is {\em $r$-reversal-bounded} if, along the transitions of the run, the factors $\lambda_i$ switch between positive and negative values at most $r$ times, for any $i$. Counter $x_i$ is {\em reversal-bounded} (shortly, \emph{r.b.}) if there is an $r$ such that, on every accepting run of $\M$, $x_i$ is $r$-reversal-bounded. $\M$ is a reversal-bounded Dense-choice Counter Machine, denoted by \emph{r.b. DCM}, if every counter in $\M$ is reversal-bounded. 

A counter which is not necessarily reversal-bounded is called a \emph{free} counter.\\

In this model, one can effectively check at runtime whether a counter is $r$-reversal-bounded, by making the control state check when transitions are incrementing ($\lambda_i > 0$) or decrementing ($\lambda_i<0$) the counter $x_i$. Thus, one can use additional control states in order to remember each reversal and crash if the number of reversals exceeds $r$.

\medskip
Like in the case of discrete counters, one can always assume that $r=1$; indeed, each sequence of "increments, then decrements" can be simulated by a $1$-r.b. counter, and thus a counter doing $r$ reversals can be simulated by $r$ $1$-r.b. counters.\\

From \cite{DenseCounter}, we know that the binary reachability of a reversal-bounded DCM with one free counter can be defined in a decidable logic: the logic of {\em mixed formulae}, 
which is equivalent to the well-known $\mathrm{FO}(\R,\Z,+,<)$. Since the syntactical details of this logic are not relevant for now, their presentation is postponed to section \ref{sec:formulae}. 
The above decidability result is stated in \cite{DenseCounter} as follows:

\begin{proposition}\label{prp:RB}
The binary reachability of a DCM with one free 0-testable counter and a finite number of reversal-bounded $k$-testable counters is definable by a mixed formula, for any $k \geq 0$.
\end{proposition}

\medskip

However, we have extended the guards of DCM to be able to test a counter against any given integer constant: we proved in Proposition \ref{prp:k-testable} that this is not more powerful in the general case, but this is far from obvious when we consider r.b. DCM. Indeed, the proof of Proposition \ref{prp:k-testable} uses an encoding which does not preserve reversal-boundedness. 

We now prove, as Theorem \ref{thm:k-test}, that this extension is actually not more powerful either in the case of r.b. DCM, provided we can use many more counters. Before that, a few more technical definitions are needed.\\

Define, for any real number $x$, $fr(x)=0$ if $\lfloor x\rfloor = x$ (i.e., $x$ is an integer), else $fr(x)=1/2$. 
Given a finite set $S$ (the control states of $\M$) and an integer $k>0$, let $S' = S \times (\{0, \dots, k\}\times \{0, 1/2\})^n$.
A DCM $\M'=\langle S',T' \rangle$ with $n$ $0$-testable counters is called a 
{\em finite-test DCM}. 

A configuration $\langle \ll s,(d_1,f_1), \dots, (d_n,f_n) \gg, \x\rangle$ of $\M'$ is {\em consistent} if, 
for every $1\le i \le n$, either $(x_i\le k \land d_i=\lfloor x_i\rfloor \ \land \ f_i=fr(x_i))$ or $(x_i>k \land d_i=k \ \land \ f_i=1/2)$ holds.
Hence, in a consistent configuration, a test of a counter against a constant $j \le k$ gives the same result as a test 
against the $d$ and $f$ components of the state.

In general, $\M'$ may also reach non-consistent configurations. A run of $\M'$ is {\em consistent} if it goes through consistent configurations only. \\


Now, we need a technical lemma showing that, for any $k$-DCM, we can build an equivalent finite-test DCM (the notion of equivalence used here is detailed in three requirements). This result will be used as a basis for constructions in the proofs of Lemma \ref{lem:k-rb} and Proposition \ref{boundedCounter}.
\begin{lemma}\label{lem:finiteState}
Let $\M=\langle S,T \rangle$  be a DCM, with $n$ $k$-testable counters $x$, with $k>0$. 
Then there exists a finite-test-DCM $\M'=\langle S',T' \rangle$, with $n$ $0$-testable counters
such that:
\begin{enumerate}
\item if counter $x_i$ of $\M$ is reversal-bounded then counter $x_i$ of $\M'$ is also reversal-bounded 
\item every run of $\M$ is also a run of $\M'$, i.e. for every run of $\M$ of length $l\ge 0$: 
\[\langle s^1, \x^1 \rangle \rightarrow_\M^l \langle s^l, \x^l \rangle,\] 
there exists a run of length $l$ for $\M'$: 
\[\langle \ll s^1,(d^1_1,f^1_1), \dots, (d^1_n,f^1_n) \gg,\x^1\rangle \rightarrow_{\M'}^l  \langle \ll s^l,(d^l_1,f^l_1), \dots, (d^l_n,f^l_n) \gg, \x^l\rangle\] 
\item a consistent run of $\M'$ is also a run of $\M$, i.e.  
for every consistent run of $\M'$ of length $l \ge 0$: 
\[\langle \ll s^1,(d^1_1,f^1_1), \dots, (d^1_n,f^1_n) \gg,\x^1\rangle \rightarrow_{\M'}^l  \langle \ll s^l,(d^l_1,f^l_1), \dots, (d^l_n,f^l_n) \gg, \x^l\rangle,\]
there exists a run of length $l$ for $\M$ of the form: 
\[\langle s^1, \x^1 \rangle \rightarrow_\M^l \langle s^l, \x^l \rangle\]
\end{enumerate}
\end{lemma}
\begin{proof}

The idea of this proof is to build $\M'$ to mimic the behaviour of $\M$, by reflecting the possible variations of a counter value into its finite state control. 
For simplicity, we consider only the case $n=1$, but the proof can easily be generalized to any number of counters. 
Therefore, a state of $S'$ is a triple $\ll s, d,f\gg$, where $s\in S$, $d$ is an integer in $0,\dots, k$ and $f$ is either 0 or 1/2.
Component $d$ is used as a discrete counter from 0 up to $k$, intended to represent  $\lfloor x_1 \rfloor$. 
Component $f$ is intended to represent the fractional part of $x_1$: 
$f=0$ is for the case $\lfloor x_1\rfloor = x_1$, $f=1/2$ otherwise. 
The definition of $T'$ is such that  
all tests of counter $x$ against a constant $0<j\leq k$ are eliminated and replaced by finite-state tests on $d$ and $f$. 
For instance, a test $x>j$ is replaced by a test $d>j \lor (d = j \land f=1/2)$. Only tests against 0 are replicated in $T'$.

Formally, $T'$ is defined as follows. \\

Let $(s,({\bf g},\boldsymbol{\lambda},{\bf a}),s') \in T$. Define $g'_1$ to be $true$ if $g_1$ is either $true$, $x_1<j$, $x_1=j$, or $x_1>j$, for every $j>0$; otherwise, define $g'_1$ 
to be $x_1=0$ or $x_1>0$ if $g_1$ is, respectively, $x_1=0$ or $x_1>0$.
Hence, $g'_1$ is obtained from $g_1$ by eliminating all tests against a constant $j>0$.
For every $d\in \{0,\dots, k\}, f\in \{0,1/2\}$, if one of the following conditions holds: 
\begin{itemize}
\item $g_1=true$, or
\item $g_1=(x_1<j)$  and $d<j$, or
\item $g_1=(x_1=j)$ and $d=j \land f=0$, or
\item $g_1=(x_1>j)$ and $d>j \lor (d=j \land f=1/2)$,
\end{itemize}
then $(\ll s,d,f\gg, (g,\lambda,a), \ll s',d',f'\gg) \in T'$ for every 
$d' \in 0,\dots, k$, $f'\in \{0,1/2\}$ such that  $ \lambda_1' = \lambda_1 \land a_1' = a_1 $ and one of the following five conditions holds: 

\begin{enumerate}
\item $\lambda_1=0 \land d'=d\land f'=f$ (stay)
\item $\lambda_1=1 \land a_1=1 \land \Bigl( (d<k \land d'=d+1\land f'=f) \lor (d=k \land d'=d \land f'=1/2) \Bigr)$ (integer increment)
\item $\lambda_1=-1 \land a_1=1 \land \biggl( (0<d<k \land d'=d-1\land f'=f) \lor \Bigl( d=k \land d'=d \land (f'=1/2 \lor f'=0) \Bigr) \biggr)$ (integer decrement) 
\item $\lambda_1=1 \land a_1=\Delta \land \Bigl( (f=0 \land d'=d\land f'=1/2) \lor (f=1/2 \land d'=d+1\land f'=0) \lor (f=1/2 \land d'=d+1\land  f'=1/2) \lor (f=1/2 \land d'=d\land f'=1/2)  \Bigr)$ (fractional increment)
\item $\lambda_1=-1 \land a_1=\Delta \land \Bigl( (f=0 \land d>0\land d'=d-1\land f'=1/2) \lor (f=1/2 \land d>0 \land d'=d-1 \land  f'=1/2) \lor (f=1/2 \land d'=d \land  f'=0) \lor (f=1/2 \land d'=d \land f'=1/2)  \Bigr)$ (fractional decrement)
\end{enumerate}
\medskip

Notice that when in cases (4) and (5) more than one alternative may hold (i.e., the disjunctions between parentheses), which correspond to nondeterministic choices of $\M'$. 
Also, in case (5) (fractional decrement), it is implicit that if $f=0 \land d=0$ then $\M'$ crashes, since there is no available alternative. \\

The above definition implements the elimination of tests. 

Let $(\ll s,d,f\gg, (g',\lambda,a),\ll s',d',f'\gg) \in T'$. 
If the original test $g_1$ is against a constant $j>0$, then $g'_1$ only requires a test of state components $d$ and $f$, but no test of $x_1$. 
If, instead,  $g_1$ is a test against 0, then 
$g'_1$ is a test whether $x_1=0$, but the finite-state control also ``tests'' that both $d$ and $f$ are 0. Similarly, if $g_1$ is $x_1>0$, then $g'_1$ is also $x_1>0$ and $d>0 \lor (d=0 \land f=1/2)$ must hold.
This entails that if at runtime there is a test $x_1 = 0$ while $x_1=0 \land (d>0 \lor f=1/2)$, then $\M'$ crashes.\\

Now, we show that the machine we defined meets the 3 conditions stated by this lemma.

Condition (1) is immediate, since one can effectively check if a dense-choice counter is $r$-reversal-bounded, for a given $r\ge 0$, by checking when transitions are incrementing ($\lambda_i > 0$) or decrementing ($\lambda_i<0$) the counter $x_i$. Thus, one can use additional control states in order to remember each reversal and crash if the number of reversals exceeds $r$.

Condition (2) of the lemma is also obvious, since by construction, every transition in a run of $\M$ may be replicated in a run of $\M'$. 
Hence, a run of $\M$ is also a run of $\M'$, by adding suitable additional components to the state. 

Condition (3) also follows, since in a consistent configuration every test of a counter against $j>0$ (with $j\le k$) is equivalent to a finite-state test. Hence, a consistent run in $\M'$ may be replicated also in $\M$. 
\end{proof}

\medskip

Notice that, in general, a run of a finite-state DCM $\M'$ (as above) is {\em not} also a run of $\M$, since in $\M'$ there is no test against constants, which are replaced by tests on state components $d$ and $f$.
Indeed, the fractional increments/decrements of the counter may lead to a non-consistent configuration where a counter value $x_i$ is not compatible with the value of the $i$-th state component $d$ and $f$, 
e.g., $x\le j$, for some $j>0$, and on the other hand $d>j$. Therefore, the tests on $d$ and $f$ may not give the same results as a test on the actual value of $x$, and hence the run may be possible in $\M'$ but not in $\M$. \\

We now extend the result of Lemma \ref{lem:finiteState} to a full equivalence relationship, this time between r.b. $k$-DCM and r.b. DCM (not finite-state, but with about $2(k+1)$ times as much counters).

\begin{lemma} 
\label{lem:k-rb}
A DCM $\M=\langle S,T\rangle$ with one free 0-testable counter and $n$ $k$-testable 1-r.b. counters 
is equivalent to a DCM $\M^0 = \langle S^0,T^0\rangle$ with one free 0-testable counter and up to $2n+2k(n+1)$ 0-testable r.b. counters.
\end{lemma}
The idea of this proof (detailed on page \pageref{prf:lem_k-rb}, in the appendix) is the following. We first build an intermediate finite-state DCM $\M'$ like in Lemma \ref{lem:finiteState}. Then, we define $\M^0$ to have its runs split in two phases. 
The first phase simulates a run of $\M'$ on the first $n$ counters, hence using finite-state tests rather than actual tests on the counters in position 1 to $n$. 
However, during this simulation phase, $\M^0$ replicates the values stored in $x_i$ into the first $n(k+1)$ additional counters. 
The second phase verifies that the simulated run of $\M'$ is actually consistent, by checking the actual counter values 
stored in the additional counters, and crashing if, and only if, the simulated run was not consistent (e.g., verifying that if $\M^0$ entered a configuration
with $d_i=j \land f_i=1/2$, then $j<x_i< j+1$). Notice that the additional counters are still reversal-bounded. 
Hence, $\M^0$ can faithfully simulate $\M$.

The proof assumes at the beginning a few restrictions on counter behaviors and tests, which are then lifted at the end.\\

Since reversal-bounded counters can always be transformed into (a larger number of) 1-r.b. counters, we can then generalize Proposition \ref{prp:k-testable} to the case of r.b. counters, by directly extending Lemma \ref{lem:k-rb}:

\begin{theorem} \label{thm:k-test}
Reversal-bounded $k$-DCM can be encoded into reversal-bounded DCM, for any $k \geq 0$.
\end{theorem}

This theorem immediately generalizes the main result of \cite{DenseCounter}, recalled here as Proposition \ref{prp:RB}


\section{Decidability and Undecidability results}

The following table summarizes the results about DCM and their variations. The results in \textbf{\textsl{bold slanted}} characters are proved in this paper, and the others were proved (or inferable) from previous papers, namely \cite{DenseCounter} and \cite{Minsky67}. There are four possible entries in this chart: ``?" if we do not know whether the state reachability problem is decidable, ``U" if it is undecidable, ``D" if it is decidable, and ``C" if the binary reachability relation is computable and definable in a decidable logic. The ``+ r.b." (resp. ``+ $k$-test. r.b.") means that the machine is extended with a finite number of reversal-bounded dense-choice counters (resp. reversal-bounded $k$-testable dense-choice counters).


\begin{center}
 \label{chart}
\begin{tabular}{|cc||c|c||c|c|}
\hline
\multicolumn{2}{|c||}{\multirow{2}{*}{counters}} &\hspace{.15cm} \multirow{2}{*}{DCM} \hspace{.15cm}&\hspace{.15cm} bounded \hspace{.15cm} & \hspace{.15cm} DCM \hspace{.15cm} & DCM + \\
& & & $k$-DCM & + r.b. & \hspace{.15cm} $k$-test. r.b. \hspace{.15cm} \\
\hline\hline
\multirow{4}{*}{\shortstack{purely\\dense-choice}} & 1 & C & \textbf{\textsl{C}} & C & \textbf{\textsl{C}} \\
& 2 & D & ? & ? & ? \\
& 3 & ? & ? & ? & ? \\
& \hspace{.15cm} 4 \hspace{.15cm} & U & \textbf{\textsl{U}} & U & \textbf{\textsl{U}} \\
\hline
\multirow{2}{*}{\hspace{.15cm} dense-choice \hspace{.15cm}} & 1 & C & \textbf{\textsl{C}} & C & \textbf{\textsl{C}} \\
& 2 & U & U & U & U \\
\hline
\end{tabular}
\end{center}

\medskip
In the remainder of this section, we prove two of these new results; the other new results are proved in the previous section or directly inferable.
Notice that the seven open problems could be solved by only two or three proofs that subsume other results. However, the intuitions about 1 or 4 counters do not fit the case of 2 or 3 counters, and the proof techniques get even more complex when we use $k$-testable counters.

\subsection{Undecidability for bounded purely dense-choice counters}

Given a DCM $\M$, a counter $x$ of $\M$ is {\em $b$-bounded}, $b \geq 0$, if $x \le b$ along every run of $\M$.
For instance, a 1-bounded counter can assume any non-negative value up to 1.
A counter is \emph{bounded} if it is $b$-bounded for some $b\ge 0$. 

Given $b \geq 0$, if a machine $\M$ has a $b$-bounded $b$-testable dense-choice counter $x$, 
then one can assume that $\M$  must crash not only when trying to decrement $ x$ below  0, but also 
when trying to set $x > b$. Indeed, if $x$ was not bounded, $\M$ could be modified to test at each step whether $ x \le b$, crashing if this is not the case (which would force $x$ to be bounded). \\

Bounded {\em integer} counters have a finite set of possible values, which can be encoded into the control states. However, bounded {\em dense-choice} counters have an infinite set of possible values: 
a DCM with several bounded counters is a powerful model, as shown next.
In general, the state reachability problem is a simpler problem than computing the binary reachability. 
However, the following proposition shows that even for state reachability, having only four 1-bounded counters
implies undecidability.

\begin{proposition}\label{prp:undec}
The state reachability problem is undecidable for bounded purely-DCM.
\end{proposition}
\begin{proof}
We show that the state reachability problem for a DCM with four purely dense-choice 1-bounded 1-testable counters is undecidable, which entails this proposition. The result follows the lines of the proof in \cite{DenseCounter} that 4 purely dense-choice counters are enough to simulate a Minsky machine. 
The original proof was based on using two counters to store a fixed value $\delta$, chosen at the beginning of the computation. The two remaining counters are  then incremented 
or decremented only of this fixed value $\delta$: any integer value $k$ is encoded as $k\delta$. Hence, the two counters behave like two discrete counters without any restriction. 
If the 4 counters are $1$-bounded, then they can encode only up to an integer $m = \lfloor 1 / \delta\rfloor$. 
However, $m$ is unbounded, since $\delta$, selected non-determistically once at the beginning of a computation, can be chosen to be arbitrarily small: 
if $\delta$ is not small enough then the DCM will crash trying to increase one of its counters beyond 1 (for example, by resetting to the initial configuration so that $\delta$ can be chosen again until it is small enough). 
But in every halting computation of a Minsky machine, the values encoded in 
its two counters are bounded (with a bound depending on the computation). Therefore, the final state of the DCM is reachable if, and only if, the simulated Minsky machine has one halting computation.
Hence, the state reachability problem is undecidable.
\end{proof}

\subsection{Decidability with one $k$-testable counter}

Proposition \ref{prp:undec} does not rule out decidability if using less than four counters, since its proof is based on a four-counter purely-DCM. 
In particular, here we show that, for a DCM with only one counter, the binary reachability can effectively be computed even if the counter is $k$-testable. 
This extension to $k$-testability is indeed far from obvious. 

In fact, the construction of the proof of Proposition~\ref{prp:k-testable} can be applied for tests of the form $x>j$  or $x=j$, for any $j\le k$; this construction can be simulated by a sequence of $j$ unit decrements followed by a test $x>0$ or $x=0$, and then followed by $j$ unit increments to restore the original value. 
However, it cannot be applied for tests of the form $x<j$, since this would require a (non-existent) additional counter to be able to restore the original counter value. 
The proof also requires the counter to be bounded, in order to avoid an unbounded number of crossings of threshold $k$. 

Again, we use the notion of mixed formula, which we develop in Section \ref{sec:formulae}. Remember that it is a decidable logic, equivalent to $\mathrm{FO}(\R,\Z,+,<)$. Moreover, we know from Proposition \ref{prp:RB} that the binary reachability of a DCM is definable by a mixed formula.

\begin{proposition}\label{boundedCounter}
The binary reachability of a DCM with a single bounded $k$-testable counter is definable by a mixed formula, for every $k \ge 0$.
\end{proposition}
\begin{proof}
Let $\M= \langle S, T\rangle$ be a one-counter DCM, such that its only counter is $b$-bounded. Since there is only one counter, a real value $x$ is used instead of a vector $\x$ of counter values.
We prove the case $b=k$, since if $b<k$ then all tests against $j>b$ are just false, while if $b>k$ then simply $\M$ will not use the tests against $k+1, k+2, $ etc.

Let $\M'= (S',T')$ be the finite-test DCM with one free 1-testable counter, as defined by Lemma~\ref{lem:finiteState}, 
with $\ll s, d,f\gg\in S'$ for every $s\in S, d\in \{0\dots k\}, f\in \{0, 1/2\}$. \\

We claim that for every $x^0, x^1\in \R_+$ and $s^0, s^1\in S$, with $s_{\rm init}, s_{\rm final} \in S'$, 
\begin{equation}\label{M'Reach}
\langle s^0,x^0\rangle \leadsto_\M \langle s^1, x^1 \rangle
\mbox{ if, and only if, }
\langle \ll s_{\rm init},\lfloor x^0 \rfloor, fr(x^0) \gg, x^0\rangle \leadsto_{\M'} \langle \ll s_{\rm final} ,\lfloor x^1 \rfloor, fr(x^1) \gg, x^1 \rangle
\end{equation}

The main proposition follows then immediately, since relation (\ref{M'Reach}) is decidable and can be described by a mixed formula.

``Only If'': This part is guaranteed by Condition (2) of Lemma~\ref{lem:finiteState}.  

``If'' part: 
Suppose that Formula (\ref{M'Reach}) holds. We need to show that $\langle s^0,x^0\rangle \leadsto_\M \langle s^1, x^1 \rangle$. 
Condition (3) of Lemma~\ref{lem:finiteState} only applies to consistent runs, and in general runs of $\M'$ may not be consistent.
However, each fractional increment/decrement in a run of $\M'$ is chosen non-deterministically. Hence the value of $x$ can be adjusted for consistency with $d$ and $f$.
The proof of this claim requires some preliminary definitions and propositions.\\

A {\em consistent version} of a configuration $c= \langle \ll s,d, f \gg, x\rangle$ 
is any configuration $c'= \langle \ll s,d, f \gg, x'\rangle$, for some $x'\in \R_+$, which is consistent. 

We claim that for every consistent configuration $c_0= \langle \ll s,d, f \gg, x^0\rangle$ 
and for every configuration $c_1=\langle \ll s_1,d_1 ,f_1 \gg, x^1\rangle$, 
\begin{equation}\label{plus}
\begin{split}
&\mbox{if } c_0 \stackrel{g',\lambda,a}{\longrightarrow}_{\M'} c_1, 
\mbox{then there exists a consistent version } c'_1 \mbox{ of } c_1 \mbox{ such that } c_0\stackrel{g',\lambda,a}{\longrightarrow}_{\M'} c'_1,\\
&\mbox{defined by } c'_1= \langle \ll s_1,d_1, f_1 \gg, {x'^1}\rangle \mbox{, for some } x'\in\R_+. 
\end{split}
\end{equation}
If $a=1$, then $c_1$ is already consistent by definition of $\M'$. 
Hence, only a fractional increment/decrement (i.e. if $a=\Delta$ and $\lambda \neq 0$) may lead to an inconsistent configuration. \\

A special case of configuration is a {\em zero-conf}, i.e. any configuration of $\M'$ of the form $\langle \ll s,(0,0)^n\gg ,{\bf 0}\rangle$. 
We can further assume that, in a finite-test DCM $\M'$, every zero-conf is always consistent, since $\M'$ can test that every component 
of $\x$ is actually 0 (and crashes otherwise). \\

Assume first $x^0=0$ (i.e., $c_0$ is a zero-conf). 
Hence, $d_1=d$, $f_1=1/2$ and $x=\delta$ for some $\delta$, $0< \delta < 1$. 
Hence, $c_1$ is already consistent. 

Assume now $x^0>0$. Hence, $d_1$ can differ from $d$ at most by one. 

To proceed, we need an additional observation. 
For all states $\ll s,d,f \gg$ and $\ll s',d',f'\gg $ of $\M'$, for all $(g,\lambda,a)\in \Sigma$, for all 
$x\in \R_+$, $\forall \epsilon \in [-1,1]$ , 
with $0\le x+\epsilon<k+1 $, 
if  
\[\langle \ll s,d, f \gg, x\rangle\stackrel{g',\lambda,a}{\longrightarrow}_{\M'} \langle \ll s',d', f' \gg, x+\epsilon\rangle\] 
then for every $x'\in \R_+$, such that 
$0\le x'+\epsilon<k+1$ 
the same move can be repeated from $x'$:

\begin{equation}\label{anyX}
\langle \ll s,d, f \gg, x'\rangle\stackrel{g',\lambda,a}{\longrightarrow}_{\M'} \langle \ll s',d', f' \gg, x'+\epsilon\rangle  
\end{equation}

Property (\ref{anyX}) is obvious since $\M'$ can only test $x$ for zero, hence it cannot differentiate $x$ from $x'$ before the move and it may apply the same increment.

By property (\ref{anyX}), it is possible to make the same move from $c_0$ using a different increment (or decrement): 
$x$ can be increased (or decreased) by a value (larger or smaller than $\delta$, 
but always in the interval $]0,1[\,$) 
which is enough to make up the difference for making the configuration consistent. \\

Let $c_0\rightarrow_{\M'} c_1\rightarrow_{\M'} \dots \rightarrow_{\M'} c_l$ be a run of $\M'$, with $l\ge0$.
We now prove by induction on $l$ that 
if $c_0$ is consistent, then there is another run of $\M'$ denoted by 
$c_0\rightarrow_{\M'} c'_1 \rightarrow_{\M'} \dots \rightarrow_{\M'} c'_l$ where each $c'_i$ is a consistent version of $c_i$, $1\le i\le l$. 

The case $l=0$ is trivial (with $c_0=c'_0$).
Suppose $l>0$. By induction hypothesis, $c_0\rightarrow_{\M'} c'_1  \rightarrow_{\M'} \dots \rightarrow_{\M'} c'_{l-1}$, 
each $c'_{i}$ being a consistent version of $c_{i}$, $1\le i\le l-1$. 
By Property (\ref{plus}), we can find a consistent version $c'_l$ of $c_l$ such that 
$c'_{l-1}\rightarrow_{\M'} c'_l$.\\

By Condition (3) of Lemma~\ref{lem:finiteState}, every consistent run of $\M'$ is also a run of $\M$; hence the proof is completed. 
\end{proof}

The proof is immediately extendable to the case where $\M$ has also discrete reversal-bounded counters, which are in no way influenced by the above construction. 
If the reversal-bounded counters are dense-choice, though, decidability is still open. This result would need a new proof, because the techniques used for Proposition \ref{boundedCounter} and Lemma \ref{lem:k-rb} apparently cannot be combined. 

\section{Logical characterization of DCM}

\subsection{Preliminary results about Mixed Formulae}
\label{sec:formulae}
We consider here the language of mixed formulae, defined in \cite{Weispfenning99}, and adapted from Presburger arithmetic. 
The language has two sorts of variables: real variables, denoted by $x, x', x_1, \dots$  and integer variables, denoted by $y, y', y_1, \dots$; the
latter are a subsort of the former. The constants are 0 and 1, the operations are $+$ (binary), $-$ (unary), $\lfloor . \rfloor$, and
the relations are equality $=$, ordering $<$ and congruences $\equiv_d$ for every constant $d\in \N$. 
Definition \ref{def:mixed} formalizes this idea:\\

\begin{definition}
\label{def:mixed}
A \emph{mixed fomula} is inductively defined as follows.
A \emph{mixed linear expression} $E$ is defined by the following grammar, where $x$ is a real variable and $y$ is an integer variable:
$$E ::= 0 \mid 1 \mid x \mid y \mid E+E \mid E-E \mid \lfloor E \rfloor$$
A \emph{mixed linear constraint} $C$ is defined by the following grammar, where $d$ is a positive integer:
$$C::= E = E | E<E | E  \equiv_d E$$
A \emph{mixed formula} $F$ is defined by the following grammar, where $x\in \R$ and $y\in \Z$:
$$F ::= C \mid \neg F \mid F\land F \mid \exists x. F \mid \exists y. F$$
\end{definition}

The semantics of a mixed formula is like in the reals, $\lfloor r \rfloor$ being the integer part of its real argument $r$, and $r_1\equiv_d r_2$ holding if $r_1-r_2 =vd$ for some integer $v$.

Typically, one can use shorthands, such as using e.g. $3x$ for $x+x+x$, or introducing other common operators (like $\ge$), etc. 

Mixed formulae are equivalent to the well-known first-order additive theory of integers and reals $\mathrm{FO}(\R,\Z,+,<)$, 
since the floor operator $\lfloor x \rfloor = y$ can be rewritten as $\exists x_1  (0<x_1 \land x_1<1  \land x-x_1 =y)$, and 
$x_1\equiv_d x_2$ can be rewritten for a fixed $d>0$ as $\exists y  (x_1 - x_2 = \underbrace{y+\dots + y}_d)$.
However,  the main advantage of the richer syntax of mixed formulae is that it allows for quantifier elimination, which is not possible in $\mathrm{FO}(\R,\Z,+,<)$, as shown in Theorem 3.1 and Corollary 5.2 of \cite{Weispfenning99}.



\subsection{Mixed formulae are definable by reversal-bounded DCM}

It is well known that reversal-bounded discrete CM can define all Presburger formulae. 
Since Presburger logic admits effective quantifier elimination, the binary reachability of r.b. discrete CM can effectively define all Presburger relations.  
A similar result holds for r.b. DCM, using mixed formulae (and the effectiveness of quantifier elimination) instead of Presburger formulae. \\

Let ${\bf 0}$ be a vector $(0,\ldots,0)$ of size $n$. 
A quantifier-free mixed formula $F(z_1, \dots, z_n)$ of ${\cal L}$ in the free variables $z_1\ge 0, \dots, z_n\ge 0$ is {\em definable} by a DCM $\M$ with at least $n$ counters $x_1, \dots, x_n$ (and possibly more) if $\M$, starting in a given initial configuration $\langle s,{\bf 0}\rangle$, may reach all, and only, final configurations $\langle s_{fin},\x\rangle$ such that   $F(x_1/z_1, \dots, x_n/z_n)$ holds (where $x_i/z_i$ denotes a substitution of variable $z_i$ with value $x_i$). 


\begin{proposition} \label{prp:formula}
Let $F(x_1, \dots, x_n, y_1, \dots, y_p)$ be a quantifier-free mixed formula, with $x_1\ge 0, \dots, x_n\ge 0, y_1\ge 0, \dots, y_p\ge0$. 
Then $F$ is definable with a r.b. DCM.
\end{proposition}

The idea of this proof (detailed page \pageref{prf:formula}, in the appendix) involves several steps which can easily be understood. First, we assume (w.l.o.g.) that $F$ is in disjunctive normal form; then, we transform it into a union of intersections of smaller formulae of the form $E \sim 0$, with $\sim \in \{>,=,<, \equiv_d, \not\equiv_d\}$. The main idea is to encode each of these formulas $E \sim 0$ by a r.b. DCM, in which there are $n$ r.b. dense-choice counters $x$, $p$ r.b. discrete counters $y$, and possibly more r.b. counters. We provide a simple r.b. DCM encoding for each of these formulae $E \sim 0$, in which the machine accepts a run, with initial counter valuations equal to the assignment of the free variables of $F$, if and only if this assignment of variables makes the formula true.

Then, we just have to connect each r.b. DCM as follows. Each machine has a final control state, which we connect with a transition to the initial control state of another machine; both of them are in fact a (bigger) machine. For the union, we add one transition going from the final state of the first machine to the initial state of each machine encoding a component of the union. Then, each of these components is a series of machines, encoding intersections. The last component of each intersection is encoded by a machine whose final state leads to an accepting sink state (which is the final state of the overall r.b. DCM encoding $F$). Finally, this sink state is reached if and only if the formula $F$ is satisfied.\\

Since the quantifier elimination of mixed formulae is effective, and since we can encode negative variables with a sign bit in the control states, then we can directly deduce the following theorem:

\begin{theorem}\label{characDCM}
Any mixed formula can be defined by a r.b. DCM.
\end{theorem}

This theorem is actually dual to the one in \cite{DenseCounter} (cited here as Proposition \ref{prp:RB}), which states that the binary reachability of a r.b. DCM (with an additional free counter) is a mixed formula. Hence, we get an exact characterization of r.b. DCM.\\

As a matter of fact, Theorem \ref{characDCM} can be combined with other results about mixed formulae. For example, we know that the binary reachability of a flat counter automaton \cite{CJ98} or of a timed automaton \cite{Dang03} is definable by a mixed formula. Hence, we can construct a r.b. DCM which is accepting exactly the binary reachability of a given flat counter automaton or timed automaton.

\section{Conclusions and future work}
The goal of this paper is to shed a more formal light on DCM, hence clarifying their relation with (discrete) CM. 
This makes us notice that there are very simple results for CM that still hold for DCM, but require a much more difficult proof. A first extension is 
to allow dense-choice counters to be compared to an integer constant $k$, and not only to 0. 
We showed that dense-choice counters are not more powerful when they are $k$-testable, even in the case of r.b. DCM, or of DCM with a single bounded counter.

A second extension is the exact characterization of r.b. DCM with the well-known first-order additive logic of integers and reals, similarly to r.b. CM with Presburger logic. 
 
We also found results that cannot be extended from CM to DCM. For instance, restricting dense-choice counters to be bounded 
does not imply decidability.

\paragraph{\bf Future work.} 
Other existing results for CM (such as the rich properties for one-counter machines) could be extended to DCM. There are also missing items in the table of page \pageref{chart}, which do not seem to be easily inferable from known results. 
One could also study different versions of dense-choice counters, such as DCM in which tests of the form $(x=0)$ are forbidden (leading to what we would call "Dense-choice Petri Nets"). We would also like to formally compare timed automata with DCM, using languages.

\bibliographystyle{eptcs}
\bibliography{biblio}

\newpage
\appendix

\section*{Appendix: Proofs}
\label{startAppendix}

\bigskip
\noindent\textbf{Lemma \ref{lem:k-rb}.} \label{prf:lem_k-rb}
\emph{A DCM $\M=\langle S,T\rangle$ with one free 0-testable counter and $n$ $k$-testable 1-r.b. counters 
is equivalent to a DCM $\M^0 = \langle S^0,T^0\rangle$ with one free 0-testable counter and up to $2n+2k(n+1)$ 0-testable r.b. counters.}
\begin{proof}
Assume that in $\M$ all r.b. counters are actually 0-reversal 
(i.e., they do not make any reversal: the counter can never ``come back'' to previous values), that all r.b. counters start from 0, and 
that there is no equality test against any $j>0$  (i.e., only $x>j$ or $x<j$  tests are allowed).
Finally, assume that $\M$ has no free counter, and hence has only $n$ $k$-testable 1-r.b. counters.
All these restrictions are then lifted at the end of the proof. 

Let $\M'=\langle S',T'\rangle$ be a finite-test DCM verifying Lemma~\ref{lem:finiteState}. 


\medskip
Consider now $\M^0$. 

Let $S^0= S' \times \{\scriptsize{\texttt{SIMUL}}, \scriptsize{\texttt{CHECK}}\}$. 
If $\M$ starts in a state $s_0$, with all counters initially at 0, then $\M^0$ starts in state $\ll (s_0, (0,0)^n ), \scriptsize{\texttt{SIMUL}}\gg$, with all counters initially at 0. 

\medskip
$\M^0$ works in two phases: first in {\scriptsize{\texttt{SIMUL}}} phase and then in {\scriptsize \texttt{CHECK}} phase. 
Correspondingly, $T^0$ is the union of two sets of transitions: $T_\texttt{\tiny{SIMUL}}$ and $T_\texttt{\tiny{CHECK}}$.

The {\scriptsize \texttt{SIMUL}} phase simulates a run of $\M'$ on the first $n$ counters, hence using finite-state tests rather than actual tests on the counters in position 1 to $n$. 
However, during the {\scriptsize \texttt{SIMUL}} phase, $\M^0$ replicates the values stored in $x_i$ into the first $n(k+1)$ additional counters.

The \texttt{\scriptsize{CHECK}} phase verifies that the simulated run of $\M'$ is actually consistent, by checking the actual counter values 
stored in the additional counters, and crashing if, and only if, the simulated run was not consistent (e.g., verifying that if $\M^0$ entered a configuration
with $d_i=j \land f_i=1/2$, then $j<x_i< j+1$). Hence, $\M^0$ can faithfully simulate $\M$.\\

For clarity, let $c(i,j)= n+(i-1)*(k+1)+j+1$, for every $1\le i \le n$, $0\le j \le k$. \\
Hence, $c(1,0), c(1,1), \dots, c(1,k)$ are the indexes of counters $x_{n+1}, x_{n+2}, \dots, x_{n+k+1}$, 
and $c(2,0),  \dots, c(2,k)$ are the indexes of counters $x_{n+k+2}, \dots, x_{n+2k+2}$, etc. 

\medskip
$T_\texttt{{\tiny SIMUL}}$ is defined as follows:
\begin{enumerate}

 \item For all $(s_1', ({\bf g'},{\boldsymbol \lambda},{\bf a}), s_2') \in T'$, the transition ${\normalsize (\ll s_1', \scriptsize{\texttt{SIMUL}}\gg, ({\bf g'},{\boldsymbol \lambda^0},{\bf a^0}),\ll s_2', \scriptsize{\texttt{SIMUL}} \gg)}$ 
is in $T_\texttt{\tiny{SIMUL}}$, if, for every  $i, 1\le i \le n$: 
\begin{itemize} 
\item $\lambda^0_i=\lambda_i, a^0_i=a_i$;
\item $\lambda^0_{c(i,0)}= \dots =\lambda^0_{c(i,d_i-1)}=0$ (i.e., the corresponding counters stay); 
\item if it is not the case that $s_1'$ is such that $d_i=k\land f_i=1/2$, then 
$\lambda^0_{c(i,d_i+1)}=\lambda^0_{c(i,d_i+2)}=  \lambda^0_{c(i,k+1)}= \lambda_{i}$ and $a_{c(i,d_i+1)}=a_{c(i,d_i+2)}=  a_{c(i,k+1)}= a_{i}$ 
(i.e., the corresponding counters make the same move as $x_i$);
\item if $s_1'$ is such that $d_i=k\land f_i=1/2$, then $\lambda^0_{c(i,d_i+1)}=0$. 
\end{itemize}

\item For every $s'\in S'$, $(\ll s', \scriptsize{\texttt{SIMUL}}\gg, ({\bf true},{\bf 0},{\bf a}),\ll s', \scriptsize{\texttt{CHECK}} \gg)$ is 
a move in $T_\texttt{\tiny{SIMUL}}$ for any {\bf a} (where ${\bf 0}$ (resp. ${\bf true}$) is the vector with 0 (resp. $true$) in every component). 
\item No move in $T_\texttt{\tiny{SIMUL}}$ is other than those defined in (1) and (2).
\end{enumerate}

The meaning of group (2) of moves is to make $\M^0$ enter the {\scriptsize \texttt{CHECK}} phase, 
which is intended to verify whether the finite-state tests used 
in the \texttt{\tiny{SIMUL}} phase were correct. \\

Without defining $T_\texttt{\tiny{CHECK}}$ formally, we describe how $\M^0$ can check that whenever entering a configuration with $d_i=j \land f_i=1/2$, for every $i, 1\le i \le n$ and 
for every $j$, $0\le j \le k$, then actually $j<x_i< j+1$.
The value of $x_i$ at the precise moment that such configuration was entered is still stored in $x_{c(i,j)}$. 
Since the counter $x_i$ cannot make any reversal, to make sure that the configuration at that time was consistent with $x_i$, it is enough to check whether $j\le x_{c(i,j)} <j+1$ and crash if this is not the case. 

Let $z_{i,j}$ be the value of $x_{c(i,j)}$ when $\M^0$ enters the {\scriptsize \texttt{CHECK}} phase.
$\M^0$ decrements $x_{c(i,j)}$ 
of exactly $j$ (using integer decrements). 
If the machine does not crash, then $z_{i,j}\ge j$. Then, to make sure that $z_{i,j} <j+1$ (i.e. $0\le x_{c(i,j)}<1$ for the current value of $x_{c(i,j)}$), 
$\M^0$ verifies that 
if $x_{i,j+1} =0$ 
then  $z_{i,j}= j$. Otherwise, $\M^0$ makes a fractional decrement of $x_{c(i,j)}$
and crashes if the result is different from 0. 
Hence, the only case where there exists a computation that does not crash is when there exists $\delta$, $0<\delta < 1$, such that  $x_{c(i,j)}=\delta$. 
$\M^0$ repeats this procedure for every $i, 1\le i \le n$, and for every $j$, $0\le j \le k$. Finally, $\M^0$ ends the computation. 
It should be clear that if $\M^0$ ends its computation without crashing then the original tests of $\M$ were guessed correctly by $\M^0$.\\

The restriction of not having equality tests can be lifted just by noticing that a test $x_i = k$ is replaced in $\M^0$ by a test whether $d_i=k \land f_i=0$. 
It is enough that the machine marks, in its finite control, the actual value of $f_i$ when $d_i$ becomes equal to $j$. 
If $f_i=0$ then the {\scriptsize \texttt{CHECK}} phase should only check whether $z_{i,j}=j$ rather than checking 
 whether  $j\le z_{i,j}<j+1$.\\

The restriction of having only 0-reversal counters can be eliminated by adding more $n(k+1)$ 0-testable 1-r-b. counters to $\M^0$, and extending the {\scriptsize \texttt{SIMUL}} phase
to use also these additional counters. Denote by ${\tilde c}(i,j)$ the value $n+n(k+1)+ (i-1)*(k+1)+j+1$. Each counter $x_{{\tilde c}(i, j)}$ 
makes the same move as $x_i$, with $0\le j \le k$ and $i\le i \le n$, as long as 
$x_i$ is in its increasing phase (i.e., $\lambda^0({\tilde c}(i,j))=\lambda(i)$). When the decreasing phase for $x_i$ starts, 
with $\M^0$ in a state $\ll s, d_1, \dots, d_n,f_1,\dots, f_n\gg$,
then {\scriptsize \texttt{SIMUL}} acts on the counters in the positions ${\tilde c}(i,j)$  with $\lambda^0$ and $a^0$ defined as follows:  
\begin{itemize} 
\item  
$\lambda^0_{{\tilde c}(i,d_i)}= \lambda^0_{{\tilde c}(i,d_i+1)} =\dots =\lambda^0_{{\tilde c}(i,k)}=0$ (i.e., the corresponding counters stay); 
\item if it is not the case that $s_1'$ is such that $d_i=k\land f_i=1/2$ then  
$\lambda^0_{{\tilde c}(i,0)}= \lambda^0_{{\tilde c}(i,1)}= \dots =  \lambda^0_{{\tilde c}(i,d_i-1)}=\lambda_{i}$ and $a^0_{{\tilde c}(i,0)}= a^0_{{\tilde c}(i,1)}= \dots =  a^0_{{\tilde c}(i,d_i-1)}=a_{i}$ (i.e., the corresponding counters make the same move as the $i$-th counter);
\item if $s_1'$ is such that $d_i=k\land f_i=1/2$, then also $\lambda^0_{{\tilde c}(i,k)}=0$ (i.e., the counter stays). 
\end{itemize}

The {\scriptsize \texttt{CHECK}} phase for these new counters in position ${\tilde c}(i,j)$ is exactly the same as for the previously introduced $n(k+1)$ counters in positions $c(i,j)$. \\

The restriction on all counters being initialized at 0 can also be lifted by making $\M^0$ guess at the beginning of the computation 
the correct values of each $d_i$ and $f_i$ and initializing $2n$ additional 1-r.b. counters with a copy of the first $n$ counters. 
This can be obtained by emptying each counter $x_i$, $1\le i \le n$, with integer decrements first and then with one fractional decrement, 
finally crashing if $x_i$ is not 0, so verifying if $d_i$ and $f_i$ were guessed correctly. 
At the same time, a new counter, say $x'_i$, is increased of the same amounts used to decrement $x_i$ to 0. 
When $x_i=0$ and the test is passed, $\M^0$ continues the computation as above, but using counter $x'_i$ instead of $x_i$ (since it stores exactly the original value of $x_i$) and starting in a state
with the previously guessed values of $d_i$ and $f_i$ rather than from $d_i=f_i=0$. 

Finally, the restriction of not having in $\M$ a free 0-testable counter can easily be removed, by adding a free 0-testable counter also to $\M^0$. 
$\M^0$ may also simulate the behaviour of this counter during the {\scriptsize \texttt{SIMUL}} phase, leaving it to stay during the {\scriptsize \texttt{CHECK}} phase. This does not affect any of the above constructions. 
\end{proof}

\bigskip
\bigskip
\bigskip
\noindent\textbf{Proposition \ref{prp:formula}.}
\emph{Let $F(x_1, \dots, x_n, y_1, \dots, y_p)$ be a quantifier-free mixed formula, with $x_1\ge 0, \dots, x_n\ge 0, y_1\ge 0, \dots, y_p\ge0$. 
Then $F$ is definable with a r.b. DCM.}
\begin{proof}
\label{prf:formula}

Assume (w.l.o.g.) that $F$ is in disjunctive normal form:
$$F = F_1 \lor F_2 \lor \ldots \lor F_m $$
Hence, it is the disjunction of clauses $F_i$ of the form:
$$F_i = F_{i_1} \land F_{i_2} \land \dots \land F_{i_{m_i}}$$
where each $F_{ij}$, in the free variables $x_1, \dots, x_n, y_1, \dots, y_p$, can always be reduced, by pushing negation to the relational symbol and by elementary algebraic transformations, to the form: 
$$E \sim  0$$ where 
$ \sim \in \{>,=,<, \equiv_d, \not\equiv_d\}$. For instance, if $F_{i_j}$ is $E_1 < E_2$, then one may check instead if $E_1 - E_2<0$, etc.

\medskip
Below we show that for every $F_{i_j}$, there exists a r.b. DCM $\M_{i_j}$ with r.b. dense-choice counters $x_1, \dots, x_n$ and r.b. discrete counters $y_1, \dots, y_p$ 
(and possibly more r.b. counters $x_{n+1}, \dots $ and $y_{p+1}, \dots$) that accepts when relation $F_{i_j}$ is verified on the initial values of the counters  $x_1, \dots, x_n,y_1, \dots, y_p$.

This immediately entails that, for each clause $F_i$, there exists a r.b. DCM $\M_i$ 
that accepts if $F_{i_1} \land F_{i_2} \land F_{i_{m_i}}$ is verified on the initial values of its r.b. counters $x_1, \dots, x_n,y_1, \dots, y_p$. 
In fact, since $F_i$ is the conjunction of all $F_{i_j}$, 
$\M_i$ is a r.b. DCM that first makes $m_i$ copies of counters $x_1, \dots, x_n, y_1, \dots, y_p$ and 
then simulates each $\M_{i_j}$ started on one of the copies. $\M_i$ accepts if, and only if, all $\M_{i_j}$ accept. 

Therefore, it is possible to build a r.b. DCM $\M'$ whose binary reachability describes relation $F$: 
$\M'$ starts with all counters equal to zero;  first, it makes nondetermistic increments of each counter, guessing
 a tuple of values for $x1, \dots, x_n, y_1, \dots, y_p$ such that at least one  $F_i$ (hence, also $F$) holds; 
second, it follows the computation of $\M_i$ described above, in order to verify that all guesses are correct. In order to end in a configuration in which the $n$ first counters hold the values of the $n$ variables making $F$ true, we will make copies of these counters so that we do not modify them during the verification that they have been guessed right. \\

To show that for every $i,j$ there actually exists a r.b. DCM $\M_{i_j}$ defining $F_{i_j}$, we first prove by induction on the structure of $E$ that the value of $E$ can be encoded
by a r.b. dense-choice counter for $|E|$ and a flag in the control state for the sign of $E$. 
Recall that a counter value can always be copied a fixed number of times, using the encoding of Figure \ref{fig:copy}. 

\medskip
The base steps of the induction are the cases when $E$ is 0, 1, $x_i$, $y_j$, which are obvious. 

Assume now that $E$ is $(E_1 + E_2)$, with a copy of $|E_1|$ and $|E_2|$ stored in two suitable r.b. counters, with their sign flags in the finite state control. 
We can assume that $E_1$ and $E_2$ have the same sign (e.g., if $E_1\ge 0  $ and $E_2<0$ then 
$E_1 + E_2$ can be rewritten as $E_1 - |E_2|$). 
We only need to consider the case when both are positive (if both are negative, then compute $|E_1| + |E_2|$, store the result in a r.b. counter with the sign flag being negative). 
The addition $E_1 + E_2$ can then be done using the encoding of Figure \ref{fig:add}.

\medskip
Assume now that $E$ is $(E_1 - E_2)$, again  with a copy of $|E_1|$ and $|E_2|$ stored in two suitable r.b. counters $x_1$ and $x_2$ (respectively), with sign flags. 
We only need to consider the case where both $E_1$ and $E_2$ are positive (the other cases can be easily eliminated or reduced to an application of $+$). 
We can also assume that $E_1 \ge E_2$. 
If $E_2>E_1$ then the machine will guess it and it may compute $E_2-E_1$ instead, changing the sign of the result. 
The computation of $E_1 - E_2$ can then be done using the encoding of Figure \ref{fig:minus}.
Notice that if the machine made the wrong guess that $E_1 \ge E_2$, while instead $E_2 < E_1$, then this procedure will crash (hence, the non-derministic choice has to be the correct one). 

\medskip
Finally, assume that $E$ is $\lfloor E' \rfloor$; then the automaton on Figure \ref{fig:intpart} can reach $s'$ from $s$ if and only if $y=\lfloor x \rfloor$.
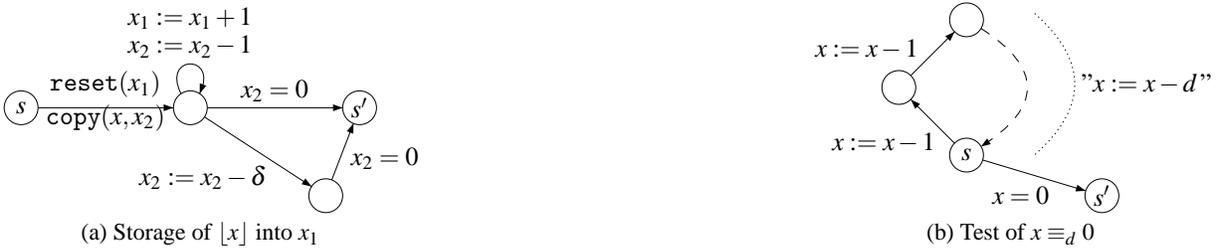
\begin{figure}[h!]
\centering
\subfloat[Storage of $\lfloor x \rfloor$ into $x_1$]{\label{fig:intpart}
\scalebox{0.9}{
    \begin{picture}(55,20)(5,0)
      \gasset{Nw=5,Nh=5}
      \node(A)(5,15){$s$}
      \node(B)(30,15){}
      \node(C)(55,15){$s'$}
      \node(D)(50,2){}
      \drawedge[ELdist=-3.4](A,B){\shortstack{$\mathtt{reset}(x_1)$\\$\mathtt{copy}(x,x_2)$}}
      \drawedge(B,C){$x_2=0$}
      \drawedge[ELside=r,ELpos=60](D,C){$x_2=0$}
      \drawedge[ELside=r,ELpos=30](B,D){$x_2:=x_2-\delta$}
      \drawloop[loopangle=90,loopdiam=4](B){\shortstack{$x_1:= x_1+1$\\$x_2:= x_2-1$}}
    \end{picture}}}
\nolinebreak
\hfill
\subfloat[Test of $x \equiv_d 0$]{\label{fig:modulo}
\scalebox{0.9}{
    \begin{picture}(55,30)(-5,0)
      \gasset{Nw=5,Nh=5}
      \node(A)(15,8){$s$}
      \node(B)(35,2){$s'$}
      \node(C)(5,18){}
      \node(D)(15,28){}
	\node[Nframe=n,Nw=0,Nh=0](X)(25,8){}
	\node[Nframe=n,Nw=0,Nh=0](Y)(25,29){}
      \drawedge[ELside=r,ELpos=45](A,B){$x=0$}
      \drawedge[ELpos=70](A,C){$x:=x-1$}
      \drawedge[ELpos=5](C,D){$x:=x-1$}
	  \drawedge[AHnb=0,ATnb=0,curvedepth=6,dash={0.2 0.5}0](Y,X){"$x:=x-d\,$"}
      \drawedge[ELpos=50,dash={1.5}0,curvedepth=9](D,A){}
    \end{picture}}}
\hfill
\caption{Encodings of ``integer part" and ``modulo"}
	\label{fig:intmod}
\end{figure}

We just showed how to encode a mixed linear expression $E$ in a r.b. DCM. To complete the proof that there is a r.b. DCM $\M_{i_j}$ accepting $F_{i_j}$, 
it is enough to show that 
there exists a r.b. DCM $\M$ which can check whether $E\sim 0$ (since $F_{i_j}$ is of this form). 
Since the value of $|E|$ is stored in a r.b. counter $x$, with a flag in the control state for the sign of $E$, then tests 
$E<0$ and $E>0$ are immediate. Of course, $E=0$ is trivial too, since it can be tested by a guard $(x=0)$. 

The two remaining cases are zero-congruences modulo an integer $d$.
The automaton on Figure \ref{fig:modulo} can reach $s'$ from $s$ if and only if $E \equiv_d 0$, for a given integer $d$ (the value of $E$ being stored into counter $x$).


To accept if $E \not\equiv_d 0$, then $\M$ first checks if  $x-\lfloor x\rfloor>0$, accepting if this is the case. If  $x-\lfloor x\rfloor=0$ then $\M$ guesses the integer constant $v \in [0,d]$ such that $E - v\equiv_d 0$. This can be computed as already explained above. \\

Thus, we gave a constructive proof that there exists a r.b. DCM defining any mixed formula.

\end{proof}

\label{endAppendix}

\end{document}